\documentclass[11pt]{article}
\usepackage{graphicx}
\renewcommand{\baselinestretch}{1.5}
\begin{document}
\title{\bf  The total energy splitting of ionic eigenstates \\ in the axial crystal fields}
\author{\bf J. Mulak$^{1}$, M. Mulak$^{2}$}
\date{{\it  $^{1}$ Trzebiatowski Institute of Low Temperature
            and Structure Research,\\
            Polish Academy of Sciences, 50--950, PO Box 1410,
            Wroclaw, Poland\\
            $^{2}$ Institute of Physics,
            Wroclaw University of Technology,\\
            Wyb. Wyspianskiego 27,
            50--370 Wroclaw, Poland}}
\maketitle
%%%%%%%%%%%%%%%%%%%%%%%%%%%%%%%%%%%%%%%%%%%%%%%%%%%%%%%%%%%%%%%%%%%%%%%%%%%
%\vspace*{1cm}
%\noindent
%{\bf Running title: The total energy splitting of ionic eigenstates ...}
%%%%%%%%%%%%%%%%%%%%%%%%%%%%%%%%%%%%%%%%%%%%%%%%%%%%%%%%%%%%%%%%%%%%%%%%%%%
\vspace*{1cm}
\noindent
{\bf Corresponding author:}\\
Prof. dr. Jacek Mulak\\
Trzebiatowski Institute of Low Temperature and
Structure Research,\\
Polish Academy of Sciences,
50--950, PO Box 1410,
Wroclaw, POLAND\\
email: Maciej.Mulak@pwr.wroc.pl\\
Tel: (+4871) 3435021, 3443206\\
Fax: (+4871) 3441029\\
\newpage
%%%%%%%%%%%%%%%%%%%%%%% Abstract %%%%%%%%%%%%%%%%%%%%%%%%%%%%%%%%%%%%%%%%%
\begin{abstract}
\noindent The relationship between the energy total splitting $\Delta E$ of the free-ion electron states in the
axial crystal-fields and the second moment of that splitting $\sigma^{2}$ is thoroughly investigated. The
non-Kramers and Kramers states with the quantum number $1\leq J \leq 8$ in the axial crystal-fields of any
multipolar composition but fixed $\sigma^{2}$ are considered. Since the crystal-field Hamiltonian ${\cal H}_{\rm
CF}$ is a superposition of the three effective multipoles various $\Delta E$ can correspond to a fixed
$\sigma^{2}$ according to the resultant combination of the independent contributions. This $\Delta E$ variation
range is the subject of the study. For the states under examination $\Delta E$ can take the values from
$2.00\sigma$ to $3.75\sigma$, whereas the difference $\Delta E_{max}- \Delta E_{min}$, except the states with
$J\leq 5/2$, amounts roughly to $\sigma$. For comparison, the one-multipolar ${\cal H}_{\rm CF}$s yield
accurately defined $\Delta E$ ranging from $2.50\sigma$ to $3.00\sigma$. The limitations of the allowed $\Delta
E$ values exclude rigorously a number of virtually possible splitting diagrams. The documentary evidence for
this restriction has been supplied in the paper collating the nominally admissible total energy splittings
$\Delta {\cal E}$ (i.e. those preserving the $\sigma^{2}$) with the $(\Delta E_{min}, \Delta E_{max})$ ranges
occurring in the actual axial crystal-fields. Although the $\Delta E$ unlike the $\sigma^{2}$ is not an
essential characteristic and depends on the reference frame orientation, it is useful to know its dispersion
range, particularly attempting to assign or verify complex electron spectra.

\end{abstract}
%%%%%%%%%%%%%%%%%%%%%%%%%%%%%%%%%%%%%%%%%%%%%%%%%%%%%%%%%%%%%%%%%%%%%%%%%%%
%%%%%%%%%%%%%%%%%%%%%%%%%%%%%%%%%%%%%%%%%%%%%%%%%%%%%%%%%%%%%%%%%%%%%%%%%%%

%--------------------------------------------------------------------------
\section*{1. Introduction}
%--------------------------------------------------------------------------
For crystal-field (CF) sublevels within any degenerate electronic state their energy center of gravity has to be
preserved, i.e their first moment always equals zero. In turn, their second moment $\sigma^{2}$, as a rotational
invariant, can serve as an appropriate measure of the real CF strength [1-5]. Comparing to $\sigma^{2}$ the
total energy splitting of the state $\Delta E$ depending on the reference frame orientation is not in fact a
marked characteristic. Nevertheless, it turns out to be really helpful to know the possible $\Delta E$ range.
Such knowledge can be valuable especially for spectroscopists trying to assign the atomic spectra, as well as to
verify the CF sublevels sequence. The above conditions for the first and second moments of the CF sublevels
confine the range of the nominally allowed total splittings $\Delta {\cal E}$ and exclude some virtual splitting
diagrams corresponding to the forbidden $\Delta {\cal E}$.

However, there appear certain additional restrictions on the total splitting $\Delta {\cal E}$. They result from
the splitting capability of the CF Hamiltonian ${\cal H}_{\rm CF}$ itself. In consequence, the range of the
experimentally observed $\Delta E$ is a narrowed subrange of the nominally admissible $\Delta {\cal E}$.

The axial three-parameter ${\cal H}_{\rm CF}$ in the tensor notation by Wybourne [6] will be considered
%%%%%%%%%%%%%%%
\begin{equation}
{\cal H}_{\rm CF}=B_{20}\;C_{0}^{(2)}+B_{40}\;C_{0}^{(4)}+B_{60}\;C_{0}^{(6)},
\end{equation}
%%%%%%%%%%%%%%%%
where $B_{k0}$ are the axial crystal-field parameters (CFPs), and $C_{0}^{(k)}$ the axial components of the
spherical tensor operator of the rank $k$.

On the one hand the energy of any CF sublevel is an algebraic sum of the three independent multipole
contributions (i.e. the $2^{(k)}$-poles for $k=2,4$ and $6$). On the other hand, due to their independence
resulting from the orthogonality of the relevant 3-j symbols [7,8] both the global second moment $\sigma^{2}$
and the partial second moments $\sigma_{k}^{2}$ are invariant with respect to the reference frame orientation.
Furthermore the multipole contributions remain additive, $\sigma^{2}=\sum_{k}\sigma_{k}^{2}$ (section 2). This
is the framework determining the $\Delta E$ variation. Therefore, the problem resolves itself into the question
-- what is the admissible variation of the total splitting $\Delta E$ of a free ion state in the axial CFs
yielding the same $\sigma^{2}$ but having different multipolar composition? To that end the maximal $\Delta
E_{max}$ and minimal $\Delta E_{min}$ values of $\Delta E$ in combined three-multipolar axial CFs will be
compared with the total splittings $\Delta E$ of the states in the individual $2^{(k)}$-pole axial fields. All
the calculated $\Delta E$ are expressed in $\sigma$ units (section 5), so they are directly related to the
well-defined experimental quantity $\sigma^{2}$. Thus, one avoids their explicite dependence on the detailed
physical parameters characterizing the charge density distribution of the surroundings and the central-ion open
shell. In consequence, the obtained results, particularly the ratio $\Delta E_{max}/\Delta E_{min}$, are free
from the typical errors. It is easy to notice that the measure of the CF strength used in this paper is based on
the produced splitting, precisely on $\sigma^{2}$. The conventional measure based on the CF strength $S$ or
$S_{k}$ [1-5, 9] in view of the scalar product character of ${\cal H}_{\rm CF}$ [4,5] is generally not adequate
(section 3). The presented results refer to the pure Russel-Saunders coupled electronic states $|\alpha
SLJ\rangle$ coming from the $^{2S+1}L$ terms. Such states have a well-defined quantum number $J$, the
degeneration $2J+1$, and $\alpha$ stands for the remaining quantum numbers needed to determine the states
completely. The calculations have been carried out for the states with $1\leq J \leq 8$. Further extension of
the analysis for the mixed states or those resulting from other couplings in the atom are feasible. To perform
this it is enough to take into account the additivity of the CF effect with respect to the constituent
$2^{(k)}$-poles in the ${\cal H}_{\rm CF}$ and to utilize the tensor transformational properties of the states
along with the standard angular momentum re-coupling techniques [7,8,10].

At the end of Discussion (section 6) a direct and instructive geometrical interpretation of the three parameter
CF interaction with the effective three component multipolar electron charge distribution of the central
paramagnetic ion in the three-dimensional coordinate system is revealed.
%--------------------------------------------------------------------------
\section*{2. The axial crystal-field}
%--------------------------------------------------------------------------
The axial CF is characterized by the  $C_{\infty v}$ point symmetry group. In consequence only three terms in
the ${\cal H}_{\rm CF}$ (Eq.(1)) from among 27 potentially possible are effective. A distinguished feature of
the axial CF is the lack of mixing of the initial substates $|\alpha SLJM_{J}\rangle$ differing in $M_{J}$. It
holds both for the individual $2^{(k)}$-poles as well as their superpositions.

The ${\cal H}_{\rm CF}$ non-zero matrix elements within the $(2J+1)$-fold $|\alpha SLJ\rangle$ state occur
exclusively on the main diagonal as the algebraic sums $\sum\limits_{k=2,4,6}(-1)^{J
-M_{J}}B_{k0}\left(\begin{array}{ccc}J&k&J\\M_{J}&0&-M_{J} \end{array} \right) \left(\langle \alpha
SLJ||C^{(k)}||\alpha SLJ\rangle\right)$, where $\left(\begin{array}{ccc}J&k&J\\M_{J}&0& -M_{J} \end{array}
\right)$ are the 3-j symbols [7,8,11], whereas $\langle \alpha SLJ||C^{(k)}||\alpha SLJ\rangle$ the multipolar
characteristics of the state [4,6]. The total splitting $\Delta E$ of the initial state, the $B_{k0}$ CFPs, as
well as sequence of the CF energy sublevels vary with the choice of the CF axis. However, the second moment of
the sublevels always remains unchanged.

Since any sublevel energy is an algebraic sum of the contributions coming from the three effective multipoles
one could rashly conclude that they can compensate either partially or even totally. The last extreme case would
lead to the resultant spherical symmetry. This reasoning is however absolutely false since it ignores the
fundamental independence of the individual additions resulting directly from the orthogonality of the
corresponding 3-j symbols. These contributions transform themselves according to various irreducible
representations of the ${\cal R}_{3}$ group. For the axial CFs the relevant orthogonality property has the form
[7,8]
%%%%%%%%%%%%%%%
\begin{equation}
\sum\limits_{M_{J}}\left(\begin{array}{ccc}J&k&J\\M_{J}&0&-M_{J}
\end{array}\right)\left(\begin{array}{ccc}J&k^{\prime}&J\\M_{J}&0&-M_{J} \end{array} \right)= \frac{\delta
(k,k^{\prime})}{2k+1},
\end{equation}
%%%%%%%%%%%%%%%%
where $\delta (k,k^{\prime})$ is the Kronecker delta. The orthogonality manifests itself in the additivity of
the second moment:
%%%%%%%%%%%%%%%
\begin{equation}
\sigma^{2}=\frac{1}{2J+1}\sum\limits_{k}B_{k0}^{2}\left(\langle \alpha SLJ||C^{(k)}||\alpha
SLJ\rangle\right)^{2}.
\end{equation}
%%%%%%%%%%%%%%%%
Such compensation can occur for an individual sublevel, but for the entire state the second moment must be
conserved. In the axial CFs there exists only one difference in the splitting diagrams of the electron states
$|\alpha SLJ\rangle$ between the non-Kramers and Kramers ions. For the non-Kramers case only one singlet
$|M_{J}=0\rangle$ appears, whereas in both cases all the remaining sublevels are always the doublets $|\pm
M_{J}\rangle$. This is why the non-Kramers state $|J\rangle$ splits into J doublets and one singlet, whereas the
Kramers state $|J\rangle$ into even number $2(n+1)$ of doublets for $J=(4n+3)/2$, or into odd number of $(2n+1)$
doublets for $J=(4n+1)/2$ $(n=0,1,2,\ldots)$. The axial CFs, more precisely the effective axial CFs, can occur
for more complex central ion coordinations due to the compensation of the off-diagonal terms in the relevant
${\cal H}_{\rm CF}$s. As an example can serve here the CF of the Archimedean antiprism symmetry ($D_{4d}\equiv
\bar{8}2m$) [11, 12]. Such approximate antiprism symmetry is met among the uranium $(4+)$ compounds, in
particular with $O^{2-}$ or $F^{-}$ ligands [12-14]. It is worth noticing, that the hexagonal as well as
trigonal ($\bar{6}$) CFs are effectively axial for $d$-electrons.

\vspace*{0.5cm}
%--------------------------------------------------------------------------------------------------------
\section*{3. The crystal-field strength}
%--------------------------------------------------------------------------------------------------------
In order to compare CF splittings produced by various multipoles either individually or collectively, a
universal accurate measure of the CF strength independent of the rank $k$ is deserved. Such a right measure
provides the second moment of the CF sublevels within the initial state $|\alpha SLJ\rangle$. It can be defined
by two equivalent ways [1-5]:

%%%%%%%%%%%%%%%
\begin{equation}
\sigma^{2}\left(|\alpha SLJ\rangle\right)=\frac{1}{2J+1}\sum\limits_{n} \left[ E_{n}-\bar{E}\left(|\alpha
SLJ\rangle\right)\right]^{2}=\frac{1}{2J+1}\sum\limits_{k}S_{k}^{2} \left(\langle \alpha SLJ||C^{(k)}||\alpha
SLJ\rangle\right)^{2},
\end{equation}
%%%%%%%%%%%%%%%%

where the center of gravity of the Stark levels within the state $|\alpha SLJ\rangle$ is given by
$\bar{E}\left(|\alpha SLJ\rangle\right)=\frac{1}{2J+1}\sum\limits_{n}E_{n}$, $E_{n}$ is the energy of the
$|n\rangle$ sublevel, and $S_{k}$ is the conventional CF strength of the $2^{k}$-pole [1-5,9] which here equals
$\frac{1}{\sqrt{2k+1}}B_{k0}$. Finally, the dimensionless scalar  $\langle \alpha SLJ||C^{(k)}||\alpha
SLJ\rangle=A_{k}\left(|\alpha SLJ\rangle\right)$ [4-6] dependent only on the angle distribution of the electron
density of the state $|\alpha SLJ\rangle$ and reflects its $2^{k}$-pole type aspherity.

Unfortunately, due to the crucial relationship $\sigma^{2}=\frac{1}{2J+1}\sum\limits_{k}S_{k}^{2}A_{k}^{2}$, the
second moment is the CF splitting measure that depends not only on the charge distribution surrounding the
central ion (represented by the $S_{k}$ or CFPs), but simultaneously on the intrinsic characteristic of the
state being affected by this CF impact (represented by $A_{k}$). The conventional measure of the CF strength
[1-3,9] $S^{2}=\sum\limits_{k}\frac{1}{2k+1}S_{k}^{2}$, which for the axial CFPs resolves itself into the
relation $S^{2}=\frac{1}{5}B_{20}^{2}+\frac{1}{9}B_{40}^{2}+\frac{1}{13}B_{60}^{2}$, is generally not useful
[4,5] because it ignores the scalar-product nature of Eqs (3) or (4). In the presented approach, having in mind
the above limitation, the CFs have the same strength if they produce the splittings with the same second moment.

Fortunately, the CF effect of any three parameter axial ${\cal H}_{\rm CF}$ ($k=2,4,6$) can be analyzed
conveniently in a three-dimensional coordinate system ($R,\theta, \phi$), which leads to an instructive
geometrical interpretation (section 6). Taking the radius of the sphere $R=\sigma$, where
$\sigma^{2}=\frac{1}{2J+1}\sum\limits_{k=2,4,6}\frac{1}{2k+1}B_{k0}^{2}A_{k}^{2}$, the energy of the sublevels
$|JM\rangle$ can be expressed in the form (in $\sigma$ units):
%%%%%%%%%%%%%%%
\begin{eqnarray}
E_{JM}&=&(-1)^{J-M_{J}}\sqrt{5(2J+1)}\left(\begin{array}{ccc}J&2&J\\M_{J}&0&-M_{J}
\end{array}\right)\sin\theta\cos\phi  \nonumber \\
&&+\;(-1)^{J-M_{J}}3\sqrt{2J+1}\left(\begin{array}{ccc}J&4&J\\M_{J}&0&-M_{J}
\end{array}\right)\sin\theta\sin\phi \nonumber \\
&&+\;(-1)^{J-M_{J}}\sqrt{13(2J+1)}\left(\begin{array}{ccc}J&6&J\\M_{J}&0&-M_{J}
\end{array}\right)cos\theta \;\;,
\end{eqnarray}
%%%%%%%%%%%%%%%%
where the coordinates $0\leq \theta < \pi$ and $0\leq \phi < 2\pi$ define the multipolar superposition in the
${\cal H}_{\rm CF}$. The first term stands for the $2^{2}$-pole (i.e. the quadrupole), the second for the
$2^{4}$-pole, and the third one for the $2^{6}$-pole, respectively. Mapping the whole variation range of the
$\theta$ and $\phi$ coordinates the upper and lower limits of the dominating absolute differences among the
sublevels energies at each ($\theta ,\phi$) point can be numerically calculated. In this way we find the
physically admissible maximum ($\Delta E_{max}$) and minimum ($\Delta E_{min}$) of the total splitting of the
$|\alpha SLJ\rangle$ state in the axial CFs yielding the same $\sigma^{2}$.

It is also worth noticing that the presented approach avoids the direct dependencies upon $B_{k0}$ and $A_{k}$
values. In consequence, the results referring to the energy span of the electron states in the axial CFs are not
burdened with typical errors introduced by $B_{k0}$ and $A_{k}$.

%--------------------------------------------------------------------------------------------------------
\section*{4. The second moment of the sublevels within the initial $|\alpha SLJ\rangle$ state
and its total nominally admissible CF splitting $\Delta {\cal E}$}
%--------------------------------------------------------------------------------------------------------
The constant second moment of the CF sublevels, $\sigma^{2}$, excessively narrows the range of the nominally
allowed total splitting $\Delta {\cal E}$ of the initial state $|\alpha SLJ\rangle$. That limitation is
different for the non-Kramers and Kramers ions. Let us also remind that not all the nominally admissible
splitting diagrams can actually occur for ions in the real CFs due to a limited splitting capability of the
relevant ${\cal H}_{\rm CF}$ (section 5).

Several hypothetical model energy diagrams in the axial CFs are compiled in Tables 1 and 2 to compare them with
the actual splittings. The data refer to fifteen $J$ values from 1 to 8. Table 1 encloses eight diagrams for the
non-Kramers ions, whereas Table 2 seven diagrams for the Kramers systems. The schemes present the sublevels
degeneration (in the parentheses), their energies expressed by the $\Delta {\cal E}$ fraction, and the relevant
expressions for the $\Delta {\cal E}$ as a function of $J$ with their corresponding values. Changing
consistently the sign in all the three $B_{k0}$ leads to the upside-down splitting diagrams. Comparing Tables 1
and 2 one can notice, as expected, that the total nominal splittings $\Delta {\cal E}$ for the non-Kramers ions
somewhat exceed those for the Kramers ones with $J$ numbers close to each other. By way of example, $\Delta
{\cal E}_{max}$ of the $|J=8\rangle$ state amounts to $5.0497\sigma$ but only $4.0000\sigma$ for the Kramers
$|J=15/2\rangle$ state. From Tables 1 and 2 results clearly the obvious inequality $\Delta {\cal E} \geq
2\sigma$. The critical equality $\Delta {\cal E} = 2\sigma$ is achieved only for the symmetric dichotomous
splitting when half of the sublevels have energy $\Delta {\cal E} /2$ and the second half $(-\Delta {\cal E}
/2)$ (see e.g. diagram 2 in Table 2). Quite instructively behave also the limits of the $\Delta {\cal E} (J)$
for $J \rightarrow \infty$. And so, on diagrams 2, 3 and 4 in Table 1, as well as on diagrams 2, 3, 4 and 5 in
Table 2 the $\Delta {\cal E}$ tends to $2\sigma$. In contrary, it tends to infinity on diagrams 1, 6, 7 and 8 in
Table 1, as well as on diagrams 1 and 7 in Table 2. The relationships $\Delta {\cal E} (J)$ for the two
homogenous diagrams (5 in Table 1 and 6 in Table 2) differ but in both cases they tend to the same limit
$3.4641\sigma$. Therefore, a question arises -- can this value lie beyond the allowed range in the actual axial
CFs? Anticipating the further analysis let us notice that such possibility could happen for sufficiently large
$J$ if the $\Delta E_{min}$, i.e. the lower limit of the admissible splittings in the actual axial CFs, exceeded
$3.4641\sigma$. For $J=8$ $\Delta E_{min}$ merely amounts to $2.4532\sigma$. Then, the homogenous sublevels
pattern would be forbidden as not respecting the condition for $\sigma^{2}$.

In the next section we compare the nominally possible $\Delta {\cal E}$ with the actual $\Delta E$ admissible in
the axial CFs. It will allow us to estimate the scale of the restrictions imposed on the $\Delta E$, as well as
to exclude some types of the splitting diagrams.
%--------------------------------------------------------------------------------------------------------
\section*{5. Results}
%--------------------------------------------------------------------------------------------------------
Table 3 comprises the physically admissible total splitting intervals $(\Delta E_{min}, \Delta E_{max})$ of the
free-ion electron states with $1\leq J\leq 8$ for all possible superpositions of the three axial $2^{k}$-poles
in the ${\cal H}_{\rm CF}$ yielding the splitting of the constant second moment $\sigma^{2}$. The upper and
lower limits of $\Delta E$ have been calculated using Eq.(5) for the CF sublevels energy $E_{JM}(\theta, \phi)$
in the axial fields within the three-dimensional frame based on the three $2^{k}$-pole contributions
$(k=2,4,6)$. In this method the whole digitized range of the angles $0\leq \theta < \pi$ and $0\leq \phi < 2\pi$
defining such multipolar composition of the superpositions is numerically swept up with the accuracy of $\;\pi
\cdot 10^{-4}$. The evaluation of the upper and lower limits has been facilitated by a short Fortran programme.
The $\Delta E_{min}$ calculations are somewhat more time-consuming comparing to  $\Delta E_{max}$ because of
their more complex character (we are looking for the global minimum of the local maxima). In fact, the $\Delta
E_{max}$ has exclusively global character and can be easily found also analytically.

For comparison, the left side of Table 3 presents the $\Delta E$ obtained for the pure individual CF multipoles
producing the splittings of the same $\sigma^{2}$. Let us here remind that the $\Delta E = 2\sigma$ determines
the absolute minimum of the total splitting. Roughly, the values of $\Delta E$ for the three individual
multipoles acting separately are close to each other for the fixed $J$ and vary from $2\sigma$ to $3\sigma$ when
$J$ increases. These are rather moderate values somewhat lower than those characteristic for the homogenous
diagrams (5 in Table 1 and 6 in Table 2). In consequence, the relevant schemes are slightly compressed with
respect to the homogenous ones.

A good example of the real splitting resembling the nominal homogenous diagram is that for the system $J=7/2$
and $k=2$, where $\Delta E = 2.6183\sigma$ vs. $\Delta {\cal E} = 2.6833\sigma$ on diagram 6 in Table 2. This
analogy is confirmed by the sequence of the corresponding CF sublevels: $(2)(-0.583), (2)(-0.084), (2)(+0.251),
 \linebreak (2)(+0.417)$, where the first bracket specifies the degeneration of the sublevel, whereas the second
its energy expressed as a $\Delta E$ fraction. The splitting of clearly asymmetrical structure occurs in the
system $J=7/2$, $k=6$ with $\Delta E = 2.4373\sigma$ vs. $\Delta {\cal E} = 2.3094\sigma$ on diagram 7 in Table
2 with the following sublevels: $(2)(-0.644), (2)(+0.071), (2+2)(+0.356)$. In turn, the splitting of distinctly
dichotomous character is met for $J=5$, $k=4$ with $\Delta E = 2.3531\sigma$ vs. $\Delta {\cal E} =
2.1409\sigma$ on diagram 4 in Table 1, with characteristic sequence of the sublevels: $(2+2)(-0.500),
(2)(+0.085), (2)(+0.335), (2+1)(+0.500)$. At last, the highest $\Delta E = 3.0461\sigma$ found for $J=3$, $k=6$
(Table 3) refers to the sequence of the sublevels: $(1)(-0.571), (2)(-0.172), (2)(+0.028),(2)(+0.429)$, which
resembles the homogenous distribution (diagram 5 in Table 1).

Nevertheless, the problem of the main interest is the resultant effect of the three component $2^{k}$-poles on
the states with different $J$. It manifests itself by the allowed minimal and maximal $\Delta E$ values (Table
3). The span of the range $(\Delta E_{min}, \Delta E_{max})$, apart from the trivial cases for $J\leq 5/2$, is
roughly equal to $\sigma$ and slowly rises with increase in $J$. For the non-Kramers ions it attains slightly
higher magnitudes. For $J=8$ and $15/2$, the difference $\Delta E_{max} - \Delta E_{min}$ amounts to
$1.2918\sigma$ and $1.1859\sigma$, respectively. When $J$ rises $\Delta E_{max}$ increases somewhat faster then
$\Delta E_{min}$, and consequently the dependencies of $(\Delta E_{max} - \Delta E_{min})$ on $J$ for the
non-Kramers and Kramers ions become similar. Worthy noticing is a distinguished position of the states with
$J=4$. Apart from the $A_{k}$ factors related to their intrinsic genealogy these states distinguish themselves
by the largest $(\Delta E_{max} - \Delta E_{min})$ difference (for $J\leq 8$), amounting to $1.4223\sigma$.
Apparently, the most favorable conditions for collective interaction of the component mulipoles must there
occur. In the next section we collate the results gathered in Tables 1 and 2 with those from Table 3. It enables
us to narrow the class of the nominally allowed CF splittings (preserving the $\sigma^{2}$) to the class of the
actual splittings in the axial CFs.
%--------------------------------------------------------------------------------------------------------
\section*{6. Discussion}
%--------------------------------------------------------------------------------------------------------
The maximal and minimal values of the nominally allowed $\Delta {\cal E}$ are achieved in the axial CFs only for
$J\leq 7/2$. It can be directly verified comparing the diagrams 1 and 2 in Tables 1 and 2 with the magnitudes of
$\Delta E_{min}$ and $\Delta E_{max}$ in Table 3. Beginning from $J=4$ neither the upper limit nor the lower
limit of the $\Delta {\cal E} $ are attainable. A decisive criterion whether a hypothetical total energy
splitting $\Delta {\cal E}$ (nominally allowed) can really occur is the natural restriction $\Delta E_{min}\leq
\Delta {\cal E} \leq \Delta E_{max}$, for the specified $J$. From among fifteen diagrams considered in Tables 1
and 2 only four are admissible without restraint. Obviously, for the pure individual CF multipoles the relevant
$\Delta E$ always lies within the range ($\Delta E_{min},\Delta E_{max})$. The values of $J$ that disqualify the
model splittings presented in Tables 1 and 2 (breaking the above inequalities) are compiled in Table 4.
Therefore, some diagrams can be a priori rejected as physically unrealistic in the axial CFs. As is also seen,
rather homogenous or close to them splitting diagrams are preferred.

There is an instructive geometrical interpretation of the CF sublevels energy as well as the conditions of their
degeneration in the axial CFs. Let us start considering Eq.(5). This energy can be written in the following form
%%%%%%%%%%%%%%%
\begin{equation}
E_{JM}=a_{JM}\;x+b_{JM}\;y+c_{JM}\;z\;\;,
\end{equation}
%%%%%%%%%%%%%%%%
where $\;a_{JM}=(-1)^{J-M_{J}}\sqrt{5(2J+1)}\left(\begin{array}{ccc}J&2&J\\M_{J}&0&-M_{J}
\end{array}\right)$, $\;b_{JM}=(-1)^{J-M_{J}}3\sqrt{2J+1}\left(\begin{array}{ccc}J&4&J\\M_{J}&0&-M_{J}
\end{array}\right)$, and $\;c_{JM}=(-1)^{J-M_{J}}\sqrt{13(2J+1)}\left(\begin{array}{ccc}J&6&J\\M_{J}&0&-M_{J}
\end{array}\right)$ are the components of a vector associated with the coordinate frame orientation, whereas $x$, $y$,
and  $z$ stand for the $\vec{\sigma}=(\sigma_{2},\sigma_{4}, \sigma_{6})$ vector components in its partition
with respect to the three $2^{k}$-poles. This is a general equation for a plane being normal to the vector
$(a_{JM},b_{JM},c_{JM})$ and distant by $E_{JM}(a_{JM}^{2}+b_{JM}^{2}+c_{JM}^{2})^{-1/2}$ from the reference
frame origin. In this geometrical representation for a constant $\sigma^{2}$ refers to the fact that the whole
space is reduced to the sphere $R=\sigma$. Thus, only those of the $B_{k0}\cdot A_{k}$ parameters (see Eq.(3))
which correspond to the intersection circles of the plane (Eq.(6)) with the sphere $R=\sigma$ fulfil the
required conditions. Degeneration of the states occurs when such circles have a common point. The state of zero
energy is represented by the plane passing through the $(0,0,0)$ point, therefore only the parameters
corresponding to the great circles of the sphere $R=\sigma$ are then involved. Solely two linearly independent
states with $E=0$ are possible. All the remaining states with $E=0$ should be represented by the planes
belonging to the pencil whose axis is determined by the two intersection points of the two circles. However,
among the reviewed states for $2\leq J\leq 8$ no linear dependence of the vectors $(a_{JM},b_{JM},c_{JM})$
occurs and, in consequence, at most two-fold degeneration for $E=0$ in the axial CFs can appear. The energy of
each sublevel $E_{JM}$ is equal to the product of the distance of the relevant plane from the $(0,0,0)$ point
and $(a_{JM}^{2}+b_{JM}^{2}+c_{JM}^{2})^{1/2}$. This normalizing factor (in the plane equation) is bigger than 1
for almost all the states $|J, M_{J}\rangle$ in the axial CFs. The exception is the $|1, \pm 1\rangle$ state for
which it amounts to $1/\sqrt{2}$, as well as the states $|3/2, \pm 3/2\rangle$, and $|3/2, \pm 1/2\rangle$ when
it equals 1. It means that all the states can reach $|E|>\sigma$. From the basic reasons it is impossible in the
case of the doublet within the singlet-doublet system of the state $|J=1\rangle$. Within the whole examined
range of the $J$ number the maximal value of the normalizing factor was found to be 2.6975 for the state $|8,
\pm 8\rangle$. The values $|E|<\sigma$ can be achieved freely since the angle between the vectors
$(a_{JM},b_{JM},c_{JM})$ and $(\sigma_{2},\sigma_{4},\sigma_{6})$ forming the scalar product can vary from $0$
to $\pi$ through $\pi/2$ what excludes any limitation for the $|E|$ from the bottom.

The presented paper is devoted to the splitting capability of the axial CFs. Their ${\cal H}_{\rm CF}$s
interaction matrices have the non-zero elements only along the main diagonal. An intriguing question arises --
how the total splitting problem looks in the case of general ${\cal H}_{\rm CF}$? What are then the admissible
ranges of the total splitting of the electron states in various CFs producing splittings of the same second
moment $\sigma^{2}$? As staring point of such further analysis may serve the expression for the second moment
$\sigma^{2}=\frac{1}{2J+1}\sum\limits_{k}S_{k}^{2}A_{k}^{2}$. It turns out that neither the rank of the $B_{kq}$
CFPs (index $k$) nor their type (index $q$) have any significance from the $\sigma^{2}$ viewpoint. Therefore the
$\sigma^{2}$ is fully determined exclusively by the sum $\sum\limits_{k}S_{k}^{2}A_{k}^{2}$.
%--------------------------------------------------------------------------------------------------------
\subsection*{Acknowledgements}
Thanks are due to Dr. P. Scharoch for his assistance in Fortran numerical calculations.
%--------------------------------------------------------------------------------------------------------

%--------------------------------------------------------------------------
% Table 1
%--------------------------------------------------------------------------
\renewcommand{\baselinestretch}{1}
\clearpage
\begin{small}
\begin{table*}[htbp]
\begin{center}
\caption{The nominal total energy span $\Delta {\cal E}$ in hypothetical splitting diagrams of $|\alpha
SLJ\rangle$ states of non-Kramers ions in axial crystal-fields (degeneration of the CF sublevels is given in the
round parentheses, their energy as a fraction of $\Delta {\cal E}$, $\Delta {\cal E} (J)$ values are given in
$\sigma$)}
%\label{tab}
\vspace*{0.8cm}
\begin{tabular}{||c|l|c|c||}
\hline
&  &  &  \\
No. & Splitting diagram  \hspace*{2cm} & Quantum number $J$ & Energy span  $\Delta {\cal E} [\sigma]$\\
&  &  &  \\
\hline \hline
&  &  &  \\
& maximum $\Delta {\cal E}$                          &        &   \\
1     &      $\left[\begin{array}{lcc}
                                      (2) &, & 1/3\\ (2J-2) & , & 0\\ (1) & , & -2/3
                    \end{array}\right.$  &
              $\begin{array}{c} 1 \\ 2 \\ 3 \\ 4 \\ 5 \\ 6 \\ 7 \\ 8 \\\end{array}$ &
              $\begin{array}{c}
2.1213 \\ 2.7386 \\ 3.2404 \\ 3.6741 \\ 4.0620 \\ 4.4159 \\ 4.7434 \\ 5.0497 \end{array}$ \\
&  $\Delta {\cal E}=\sigma\sqrt{\frac{3(2J+1)}{2}}$  &        &   \\
\hline
&  &  &  \\
& minimum $\Delta {\cal E}$                          &        &   \\
2     &      $\left[\begin{array}{lcc}
                                      (J+1) &, & \frac{J}{2J+1}\\ (J) & , & -\frac{J+1}{2J+1}\\
                    \end{array}\right.$  &
              $\begin{array}{c} 1 \\ 2 \\ 3 \\ 4 \\ 5 \\ 6 \\ 7 \\ 8 \\\end{array}$ &
              $\begin{array}{c}
2.1213 \\ 2.0412 \\ 2.0209 \\ 2.0124 \\ 2.0082 \\ 2.0058 \\ 2.0043 \\ 2.0034 \end{array}$ \\
&  $\Delta {\cal E}=\sigma\frac{2J+1}{\sqrt{J(J+1)}}$  &        &   \\
\hline
&  &  &  \\
& $J$ even                    &        &   \\
3     &      $\left[\begin{array}{lcc}
                                      (J) &, & 1/2\\ (1) & , & 0\\ (J) & , & -1/2
                    \end{array}\right.$  &
              $\begin{array}{c} 2 \\ 4 \\ 6 \\ 8 \\ \end{array}$ &
              $\begin{array}{c}
2.2361 \\ 2.1213 \\ 2.0819 \\ 2.0616 \\             \end{array}$ \\
&  $\Delta {\cal E}=\sigma\sqrt{\frac{2(2J+1)}{J}}$   &        &   \\
\hline
&  &  &  \\
& $J$ odd                    &        &   \\
4     &      $\left[\begin{array}{lcc}
                                      (J+1) &, & \frac{J-1}{2J}\\ (1) & , & 0\\ (J-1) & , & -\frac{J+1}{2J}
                    \end{array}\right.$  &
              $\begin{array}{c} 1 \\ 3 \\ 5 \\ 7 \\ \end{array}$ &
              $\begin{array}{c}
- \\ 2.2913 \\ 2.1409 \\ 2.0918 \\             \end{array}$ \\
%&  &  &   \\
&  $\Delta {\cal E}=\sigma\sqrt{\frac{2J(2J+1)}{J^{2}-1}}$   &        &   \\
\hline
\end{tabular}
\end{center}
\end{table*}
\clearpage
\begin{table*}[htbp]
\noindent Table 1 - cont.
\begin{center}
%\caption{The nominal total energy span ....}
%\label{tab}
\vspace*{0.3cm}
\begin{tabular}{||c|l|c|c||}
\hline
&  &  &  \\
No. & Splitting diagram & Quantum number $J$ & Energy span $\Delta {\cal E} [\sigma]$ \\
&  &  &  \\
\hline \hline
&  &  &  \\
& homogenous splitting \dag                         &        &   \\
5     &      $\left[\begin{array}{lcc}
                                      (2) &, & \frac{J}{2J+1}\\
                                       \vdots & , &\vdots\\
                                      (2) &, & -\frac{J}{2J+1}+\frac{2}{J}\\
                                      (2) &, & -\frac{J}{2J+1}+\frac{1}{J}\\
                                      (1) &, & -\frac{J}{2J+1}\\
                    \end{array}\right.$  &
              $\begin{array}{c} 1 \\ 2 \\ 3 \\ 4 \\ 5 \\ 6 \\ 7 \\ 8 \\\end{array}$ &
              $\begin{array}{c}
2.1213 \\ 2.6726 \\ 2.9122 \\ 3.0426 \\ 3.1239 \\ 3.1791 \\ 3.2189 \\ 3.2490 \end{array}$ \\
%&  &  &   \\
&  $\Delta {\cal E}=\sigma(2J+1)\sqrt{\frac{3J}{(J+1)(J^{2}+J+1)}}$  &        &   \\
\hline
&  &  &  \\
& supreme asymmetric splitting                          &        &   \\
6     &      $\left[\begin{array}{lcc}
                                      (2J) &, & \frac{1}{2J+1}\\ (1) & , & -\frac{2J}{2J+1}
                    \end{array}\right.$  &
              $\begin{array}{c} 1 \\ 2 \\ 3 \\ 4 \\ 5 \\ 6 \\ 7 \\ 8 \\\end{array}$ &
              $\begin{array}{c}
2.1213 \\ 2.5000 \\ 2.8577 \\ 3.1821 \\ 3.4785 \\ 3.7527 \\ 4.0089 \\ 4.2501 \end{array}$ \\
&  $\Delta {\cal E}=\sigma\frac{2J+1}{\sqrt{2J}}$  &        &   \\
\hline
&  &  &  \\
7     &      $\left[\begin{array}{lcc}
                                      (2J-1) &, & \frac{2}{2J+1}\\ (2) & , & -\frac{2J-1}{2J+1}
                    \end{array}\right.$  &
              $\begin{array}{c} 1 \\ 2 \\ 3 \\ 4 \\ 5 \\ 6 \\ 7 \\ 8 \\\end{array}$ &
              $\begin{array}{c}
2.1213 \\ 2.0412 \\ 2.2137 \\ 2.4054 \\ 2.5927 \\ 2.7716 \\ 2.9417 \\ 3.1038 \end{array}$ \\
&  $\Delta {\cal E}=\sigma\frac{2J+1}{\sqrt{2(2J-1)}}$  &        &   \\
\hline \hline
&  &  &  \\
8     &      $\left[\begin{array}{lcc}
                                      (2) &, & 1/2\\ (2J-3) & , & 0\\ (2) & , & -1/2
                    \end{array}\right.$  &
              $\begin{array}{c} 1 \\ 2 \\ 3 \\ 4 \\ 5 \\ 6 \\ 7 \\ 8 \\\end{array}$ &
              $\begin{array}{c}
- \\ 2.2361 \\ 2.6458 \\ 3.0000 \\ 3.3166 \\ 3.6056 \\ 3.8730 \\ 4.1231 \end{array}$ \\
&  $\Delta {\cal E}=\sigma\sqrt{2J+1}$  &        &   \\
\hline \multicolumn{4}{l}{\dag  {\small Other diagrams with the singlet not at the bottom but in different
location lead to similar $\Delta {\cal E} (J)$}}
\end{tabular}
\end{center}
\end{table*}
\end{small}

\clearpage

%--------------------------------------------------------------------------
% Table 2
%--------------------------------------------------------------------------
\renewcommand{\baselinestretch}{1}
\clearpage
\begin{small}
\begin{table*}[htbp]
\begin{center}
\caption{The nominal total energy span $\Delta {\cal E}$ in hypothetical splitting diagrams of $|\alpha
SLJ\rangle$ states of Kramers ions in axial crystal-fields (degeneration of the CF sublevels is given in the
round parentheses, their energy as a fraction of $\Delta {\cal E}$, $\Delta {\cal E} (J)$ values are given in
$\sigma$)}
%\label{tab}
\vspace*{0.8cm}
\begin{tabular}{||c|l|c|c||}
\hline
&  &  &  \\
No. & Splitting diagram  \hspace*{2cm} & Quantum number $J$ & Energy span  $\Delta {\cal E} [\sigma]$\\
&  &  &  \\
\hline \hline
&  &  &  \\
& maximum $\Delta {\cal E}$                          &        &   \\
1     &      $\left[\begin{array}{lcc}
                                      (2) &, & 1/2\\ (2J-3) & , & 0\\ (2) & , & -1/2
                    \end{array}\right.$  &
              $\begin{array}{c} 3/2 \\ 5/2 \\ 7/2 \\ 9/2 \\ 11/2 \\ 13/2 \\ 15/2 \\\end{array}$ &
              $\begin{array}{c}
2.0000 \\ 2.4495 \\ 2.8284 \\ 3.1623 \\ 3.4641 \\ 3.7417 \\ 4.0000  \end{array}$ \\
%&  &  &   \\
&  $\Delta {\cal E}=\sigma\sqrt{2J+1}$  &        &   \\
\hline
&  &  &  \\
& minimum $\Delta {\cal E}$                          &        &   \\
&  even number of doublets &  &  \\
2     &      $\left[\begin{array}{lcc}
                                      \left(\frac{2J+1}{2}\right) &, & 1/2\\ \left(\frac{2J+1}{2}\right) &, & -1/2\\
                    \end{array}\right.$  &
              $\begin{array}{c} 3/2 \\ 7/2 \\ 11/2 \\ 15/2 \\\end{array}$ &
              $\begin{array}{c}
2.0000 \\ 2.0000 \\ 2.0000 \\ 2.0000  \end{array}$ \\
%&  &  &   \\
&  $\Delta {\cal E}=2\sigma$  &        &   \\
\hline
&  &  &  \\
& minimum $\Delta {\cal E}$                    &        &   \\
&  odd number of doublets &  &  \\
3     &      $\left[\begin{array}{lcc}
                                      \left(\frac{2J+3}{2}\right) &, & \frac{2J-1}{2(2J+1)}\\
                                      \left(\frac{2J-1}{2}\right) & , & -\frac{2J+3}{2(2J+1)}\\
                    \end{array}\right.$  &
              $\begin{array}{c}5/2 \\ 9/2 \\ 13/2 \\ \end{array}$ &
              $\begin{array}{c}
2.1213 \\ 2.0413 \\ 2.0209 \\             \end{array}$ \\
%&  &  &   \\
&  $\Delta {\cal E}=\sigma\frac{2(2J+1)}{\sqrt{(2J+3)(2J-1)}}$   &        &   \\
\hline
&  &  &  \\
4     &      $\left[\begin{array}{lcc}
                                      \left(\frac{2J-1}{2}\right) &, & 1/2\\ (2) & , & 0\\
                                      \left(\frac{2J-1}{2}\right) &, & -1/2
                    \end{array}\right.$  &
              $\begin{array}{c} 5/2 \\ 9/2 \\ 13/2 \\ \end{array}$ &
              $\begin{array}{c}
2.4495 \\ 2.2361 \\ 2.1605 \\             \end{array}$ \\
%&  &  &   \\
&  $\Delta {\cal E}=2\sigma\sqrt{\frac{2J+1}{2J-1}}$   &        &   \\
\hline
&  &  &  \\
5     &      $\left[\begin{array}{lcc}
                                      \left(\frac{2J+1}{2}\right) &, & \frac{2J-3}{2(2J-1)}\\ (2) & , & 0\\
                                      \left(\frac{2J-3}{2}\right) &, & -\frac{2J+1}{2(2J-1)}
                    \end{array}\right.$  &
              $\begin{array}{c} 3/2 \\ 7/2 \\ 11/2 \\ 15/2 \\\end{array}$ &
              $\begin{array}{c}
- \\ 2.4495 \\ 2.2361 \\ 2.1605 \\             \end{array}$ \\
%&  &  &   \\
&  $\Delta {\cal E}=2\sigma\sqrt{\frac{2J-1}{2J-3}}$   &        &   \\
\hline
\end{tabular}
\end{center}
\end{table*}

\clearpage

\begin{table*}[htbp]
\noindent Table 2 - cont.
\begin{center}
%\caption{The nominal total energy span ....}
%\label{tab}
\vspace*{0.8cm}
\begin{tabular}{||c|l|c|c||}
\hline
&  &  &  \\
No. & Splitting diagram & Quantum number $J$ & Energy span $\Delta {\cal E} [\sigma]$ \\
&  &  &  \\
\hline \hline
&  &  &  \\
& homogenous splitting                          &        &   \\
6     &      $\left[\begin{array}{lcc}
                                      (2) &, & 1/2\\
                                       \vdots & , &\vdots\\
                                      (2) &, & -1/2+\frac{4}{2J-1}\\
                                      (2) &, & -1/2+\frac{2}{2J-1}\\
                                      (2) &, & -1/2\\
                    \end{array}\right.$  &
              $\begin{array}{c} 3/2 \\ 5/2 \\ 7/2 \\ 9/2 \\ 11/2 \\ 13/2 \\ 15/2 \\\end{array}$ &
              $\begin{array}{c}
2.0000 \\ 2.4495 \\ 2.6833 \\ 2.8284 \\ 2.9279 \\ 3.0000 \\ 3.0551 \end{array}$ \\
%&  &  &   \\
&  $\Delta {\cal E}=2\sigma \sqrt{\frac{3(2J-1)}{(2J+3)}}$  &        &   \\
\hline
&  &  &  \\
& supreme asymmetric splitting                          &        &   \\
7     &      $\left[\begin{array}{lcc}
                                      (2J-1) &, & \frac{2}{2J+1}\\ (2) & , & -\frac{2J-1}{2J+1}
                    \end{array}\right.$  &
              $\begin{array}{c} 3/2 \\ 5/2 \\ 7/2 \\ 9/2 \\ 11/2 \\ 13/2 \\ 15/2 \\\end{array}$ &
              $\begin{array}{c}
2.0000 \\ 2.1213 \\ 2.3094 \\ 2.5000 \\ 2.6833 \\ 2.8577 \\ 3.0237  \end{array}$ \\
%&  &  &   \\
&  $\Delta {\cal E}=\sigma\frac{2J+1}{\sqrt{2(2J-1)}}$  &        &   \\
\hline
\end{tabular}
\end{center}
\end{table*}
\end{small}

\clearpage

%--------------------------------------------------------------------------
% Table 3
%--------------------------------------------------------------------------
\renewcommand{\baselinestretch}{1.5}
\clearpage
\begin{small}
\begin{table*}[htbp]
\begin{center}
\caption{The total splittings $\Delta E$ of $|\alpha SLJ\rangle$ states in axial ${\cal H}_{\rm CF}$s yielding
constant $\sigma^{2}$. Comparison of the $(\Delta E_{min},\Delta E_{max})$ ranges for the $2^{k}$-pole
superpositions with the $\Delta E$ values for the pure component multipoles (all values are given in $\sigma$)}
%\label{tab}

\vspace*{1.2cm}

\begin{tabular}{||l|c|c|c|c|c||}
\hline
Quantum     &    \multicolumn{3}{c|}{The total energy span}       &
                 \multicolumn{2}{c||}{The range of the total energy spans }                  \\
number      &    \multicolumn{3}{c|}{ in the pure $2^{k}$-pole ${\cal H}_{\rm CF}$}
            &    \multicolumn{2}{c||}{for the multipole superpositions ${\cal H}_{\rm CF}$}  \\
\hline
%        &         &          &          &                    &                              \\
J& $k=2$  \hspace*{0.1cm} & $k=4$ \hspace*{0.1cm}   & $k=6$  \hspace*{0.1cm}  &
\hspace*{0.7cm} $\Delta E_{min}$ \hspace*{0.7cm}& $\Delta E_{max}$                     \\
\hline
1        &  2.1213 &  ---     &   ---    &     2.1213         &   2.1213               \\
3/2      &  2.0000 &  ---     &   ---    &     2.0000         &   2.0000               \\
2        &  2.3906 &  2.6725  &   ---    &     2.0413         &   2.7385               \\
5/2      &  2.4056 &  2.3148  &   ---    &     2.1213         &   2.4495               \\
3        &  2.5984 &  2.7717  &   3.0461 &     2.0208         &   3.2403               \\
7/2      &  2.6183 &  2.5074  &   2.4373 &     2.0000         &   2.8284               \\
4        &  2.7351 &  2.6154  &   2.8317 &     2.0802         &   3.5025               \\
9/2      &  2.7528 &  2.3651  &   2.5848 &     2.1820         &   3.1424               \\
5        &  2.8307 &  2.3531  &   2.6302 &     2.2186         &   3.3534               \\
11/2     &  2.8444 &  2.5544  &   2.8953 &     2.3075         &   3.3182               \\
6        &  2.9007 &  2.6944  &   2.9641 &     2.3350         &   3.4603               \\
13/2     &  2.9125 &  2.7883  &   2.8964 &     2.3710         &   3.4518               \\
7        &  2.9551 &  2.8478  &   2.8656 &     2.4081         &   3.6067               \\
15/2     &  2.9640 &  2.8812  &   2.9306 &     2.4293         &   3.6152               \\
8        &  2.9975 &  2.8957  &   2.9004 &     2.4532         &   3.7450               \\
\hline
\end{tabular}
\end{center}
\end{table*}
\end{small}
\clearpage

%--------------------------------------------------------------------------
% Table 4
%--------------------------------------------------------------------------
\renewcommand{\baselinestretch}{1}
\clearpage
\begin{small}
\begin{table*}[htbp]
\begin{center}
\caption{ The $J$ values disqualifying the model splitting diagrams in axial crystal-fields, $\Delta {\cal E} <
\Delta E_{min}$ or $\Delta {\cal E} > \Delta E_{max}$}
%\label{tab}

\vspace*{0.8cm}

\begin{tabular}{||c|c|c||}
\hline
 &    &    \\
Splitting diagram  & Table 1 (non-Kramers ions) & Table 2 (Kramers ions)\\
 &    &    \\
\hline
 &    &    \\
1         &  $J\geq 4$      & $J\geq 9/2$  \\
 &    &    \\
2         &  $J\geq 5$      & $J=7/2, 11/2, 15/2$  \\
 &    &    \\
3         &  $J\geq 6$      & $J=9/2, 13/2$  \\
 &    &    \\
4         &  $J\geq 5$      & $J=13/2$  \\
 &    &    \\
5         &  without restraint      & $J=11/2, 15/2$  \\
 &    &    \\
6         &  $J\geq 5$      & without restraint  \\
 &    &    \\
7         &  without restraint      & without restraint  \\
 &    &    \\
8         &  $J\geq 6$      & ---  \\
 &    &    \\
\hline
\end{tabular}
\end{center}
\end{table*}
\end{small}
\clearpage

%--------------------------------------------------------------------------

%--------------------------------------------------------------------------

\end{document}